\documentclass[12pt]{article}
\def\be{\begin{equation}}
\def\ee{\end{equation}}
\def\bea{\begin{eqnarray}}
\def\eea{\end{eqnarray}}
\usepackage{graphicx}

\catcode`\@=11
\def\lsim{\mathrel{\mathpalette\@versim<}}
\def\gsim{\mathrel{\mathpalette\@versim>}}
\def\@versim#1#2{\vcenter{\offinterlineskip
\ialign{$\m@th#1\hfil##\hfil$\crcr#2\crcr\sim\crcr } }}
\catcode`\@=12
\usepackage{axodraw}

\catcode`@=12
\begin{document}
\thispagestyle{empty}
\begin{flushright}
UCRHEP-T492\\
June 2010\
\end{flushright}
\vspace{0.3in}
\begin{center}
{\LARGE \bf Quark and Lepton Flavor Triality\\}
\vspace{1.5in}
{\bf Ernest Ma\\}
\vspace{0.2in}
{\sl Department of Physics and Astronomy, University of California,\\ 
Riverside, California 92521, USA\\}
\end{center}
\vspace{1.5in}
\begin{abstract}\
Motivated by the success of $A_4$ in explaining neutrino tribimaximal mixing, 
and its approximate residual $Z_3$ symmetry in the quark and charged-lepton 
sectors, the notion of {\it flavor triality} is proposed.  Under this 
hypothesis, certain processes such as $\tau^+ \to \mu^+ \mu^+ e^-$ 
and $\tau^+ \to e^+ e^+ \mu^-$ are favored, but $\tau^+ \to \mu^+ e^+ e^-$ 
and $\mu^+ \to e^+ e^+ e^-$ are disfavored.  Similarly, $B^0 \to \tau^+ e^-$ 
is favored, but $B^0 \to \tau^- e^+$ is disfavored.
\end{abstract}

\newpage
\baselineskip 24pt
\underline{\it Introduction}~:~  The observed neutrino mixing matrix is very 
close to the tribimaximal form~\cite{hps02} and is best understood in terms 
of the tetrahedral symmetry $A_4$~\cite{mr01,bmv03,m04,af05}.  The key to 
its success is the pattern of symmetry breaking with preserved 
subgroups~\cite{l07,bhl08} such that $A_4 \to Z_3$ in the charged-lepton 
sector and $A_4 \to Z_2$ in the neutrino sector, which may be 
accomplished~\cite{m10} in a renormalizable model using Higgs doublets 
and triplets, both transforming as \underline{1} and \underline{3} of $A_4$ 
alone~\cite{m06-2}. Other more complicated scenarios with additional 
auxiliary symmetries have also been 
proposed~\cite{bh05,af06,m06-1,h07,mpt07,m07,afh08,bfm08,m09,am09}.

In the quark sector, the same pattern for both $up$ and $down$ quarks as 
the charged-lepton sector is the most natural choice, in which case 
there is perfect alignment of the two mass matrices, resulting in 
no mixing at all.  Since quark mixing angles are known to be small, 
this is a good first approximation.  Thus the residual symmetry $Z_3$ 
is approximately valid in all fermion sectors, except the neutrino sector. 
The notion of {\it flavor triality} in the quark and charged-lepton sectors 
is then a useful tool for understanding if $A_4$ (or some other symmetry 
which also breaks to $Z_3$) is the correct underlying explanation of 
neutrino tribimaximal mixing.

\underline{\it Flavor Triality}~:~ In most applications of the non-Abelian 
discrete flavor symmetry $A_4$ to quark and lepton mass matrices, $A_4$ 
is broken spontaneously and softly to the residual symmetry $Z_3$ in the 
$up$ and $down$ quark and charged-lepton sectors, and to $Z_2$ in the 
neutrino sector.  The latter is actually a small perturbation, which 
is crucial for deriving the tribimaximal mixing of the neutrino mass matrix, 
but otherwise not very important in other physical processes.  In fact, 
the Lagrangian is approximately invariant under $Z_3$ in such a scenario. 
Let $\omega = \exp(2 \pi i/3) = -1/2 + i \sqrt{3}/2$ with $\omega^3=1$, 
then the quarks and charged leptons may be classified according to~\cite{m10}
\begin{eqnarray}
&& u,~d,~e \sim 1, ~~~ c,~s,~\mu \sim \omega^2, ~~~ t,~b,~\tau \sim \omega,\\ 
&& u^c,~d^c,~e^c \sim 1, ~~~ c^c,~s^c,~\mu^c \sim \omega, ~~~ t^c,~b^c,~\tau^c 
\sim \omega^2.
\end{eqnarray}
Therefore, certain quark and lepton flavor-violating processes are 
favored, but others are disfavored.  

Using Eqs.~(1) and (2), it is clear that only the following flavor-changing 
leptonic decays are favored:
\begin{equation}
\tau^+ \to \mu^+ \mu^+ e^-, ~~~~ \tau^+ \to e^+ e^+ \mu^-.
\end{equation}
In the quark sector, they are
\begin{equation}
b \to s s \bar{d}, ~~~ b \to d d \bar{s}, ~~~ t \to c c \bar{u}, ~~~~ 
t \to u u \bar{c}.
\end{equation}
In processes involving both quarks and leptons, they are
\begin{eqnarray}
&& B^0 \to \tau^+ e^-, ~\mu^+ \tau^-, ~e^+ \mu^-, ~~~ B_s^0 \to \tau^+ \mu^-, 
~\mu^+ e^-, ~e^+ \tau^-, \\
&& D^0 \to \tau^- e^+, ~e^- \mu^+, ~~~~ K^0 \to \mu^+ e^-, ~~~~ \tau^+ 
\to \mu^+ K^0, ~~~ \tau^+ \to e^+ \bar{K}^0.
\end{eqnarray}

\noindent \underline{\it Scalar Mediators}~:~ 
The mediators of flavor triality are a triplet (under $A_4$) of Higgs scalar 
doublets (under $SU(2)_L \times U(1)_Y$) which transform under $Z_3$ as
\begin{equation}
\phi_0 \sim 1, ~~~ \phi_1 \sim \omega, ~~~ \phi_2 \sim \omega^2.
\end{equation}
As for their Yukawa couplings to quarks and leptons, there are two 
existing versions.  One~\cite{mr01,m09} comes from having $l^c_{1,2,3}$ 
transforming as $\underline{1}, \underline{1}', \underline{1}''$ under 
$A_4$.  The other~\cite{m10,m06-2} uses $\underline{3}$.  In the former, 
$\phi_0$ is identified as the one Higgs doublet of the Standard Model, 
with couplings $(g/\sqrt{2} M_W)[m_e \overline{e}_L e_R + m_\mu 
\overline{\mu}_L \mu_R + m_\tau \overline{\tau}_L \tau_R]$ and similarly for 
quarks.  The leptonic interactions of $\phi_{1,2}$ are given by~\cite{mr01}
\begin{eqnarray}
{\cal L}_{int} &=& \sqrt{3 \over 2} {g \over M_W} [m_\tau 
\overline{(\nu_\mu,\mu)}_L \tau_R + m_\mu \overline{(\nu_e,e)}_L \mu_R + 
m_e \overline{(\nu_\tau,\tau)}_L e_R] \phi_1 \nonumber \\ 
&+& \sqrt{3 \over 2} {g \over M_W} [m_\tau 
\overline{(\nu_e,e)}_L \tau_R + m_\mu \overline{(\nu_\tau,\tau)}_L \mu_R + 
m_e \overline{(\nu_\mu,\mu)}_L e_R] \phi_2 + H.c., 
\end{eqnarray}
whereas those involving quarks are
\begin{eqnarray}
{\cal L}_{int} &=& \sqrt{3 \over 2} {g \over M_W} [m_b 
\overline{(c,s)}_L b_R + m_s \overline{(u,d)}_L s_R + 
m_d \overline{(t,b)}_L d_R] \phi_1 \nonumber \\ 
&+& \sqrt{3 \over 2} {g \over M_W} [m_b 
\overline{(u,d)}_L b_R + m_s \overline{(t,b)}_L s_R + 
m_d \overline{(c,s)}_L d_R] \phi_2 \nonumber \\ 
&+& \sqrt{3 \over 2} {g \over M_W} [m_t 
\overline{(c,s)}_L t_R + m_c \overline{(u,d)}_L c_R + 
m_u \overline{(t,b)}_L u_R] \tilde{\phi}_2 \nonumber \\ 
&+& \sqrt{3 \over 2} {g \over M_W} [m_t 
\overline{(u,d)}_L t_R + m_c \overline{(t,b)}_L c_R + 
m_u \overline{(c,s)}_L u_R] \tilde{\phi}_1 + H.c.,
\end{eqnarray}
where $\tilde{\phi} = i \sigma_2 \phi^*$.  Note that flavor-changing 
radiative decays such as $\mu \to e \gamma$ and $b \to s \gamma$ 
are not induced by these interactions.

The charged scalars $\phi^\pm_{1,2}$ are degenerate in mass~\cite{mr01}. 
However, the situation is more complicated~\cite{m09} for the neutral scalars. 
Since $\phi_1^0, \bar{\phi}^0_2$ transform as $\omega$ under $Z_3$, whereas 
$\phi_2^0, \bar{\phi}^0_1$ transform as $\omega^2$, neither $\phi_1^0$ 
nor $\phi_2^0$ are mass eigenstates. Rather, they are
\begin{equation}
\psi^0_{1,2} = (\phi_1^0 \pm \bar{\phi}^0_2)/\sqrt{2} \sim \omega,
\end{equation}
with $m_1 \neq m_2$.  Note also that in the Higgs potential itself, soft 
breaking of $A_4$ to $Z_3$ allows $\phi_{1,2}$ to have a mass different from 
that of $\phi_0$.  This mass is not related to electroweak symmetry 
breaking and subject only to specific phenomenological constraints, 
as will be discussed. 

In the latter version~\cite{m10,m06-2}, there is an additional Higgs doublet 
$\eta$ transforming as \underline{1} under $A_4$.  Whereas $\eta^0$ with 
$\langle \eta^0 \rangle = v_0$ couples to charged leptons according to
\begin{equation}
{1 \over 3 v_0} (m_e + m_\mu + m_\tau)(\bar{e}_L e_R + \bar{\mu}_L \mu_R + 
\bar{\tau}_L \tau_R),
\end{equation}
$\phi_0$ of Eq.~(7) with $\langle \phi_0^0 \rangle = v$ couples according to
\begin{equation}
{1 \over 3 v} [(2 m_e - m_\mu - m_\tau) \bar{e}_L e_R + 
(2 m_\mu - m_\tau - m_e) \bar{\mu}_L \mu_R + (2 m_\tau - m_e - m_\mu) 
\bar{\tau}_L \tau_R].
\end{equation}
Hence the linear combination $(v_0 \eta +  v \phi_0)/
\sqrt{v_0^2 +  v^2}$ acts as the Standard-Model Higgs doublet, 
and the orthogonal combination has enhanced couplings~\cite{l10} 
to $\bar{e}_L e_R$ and $\bar{\mu}_L \mu_R$.

The Yukawa couplings of $\phi^0_{1,2}$ to leptons are now given by
\begin{eqnarray}
&& {\phi_1^0 \over 3 v}[(2m_\tau - m_e - m_\mu) \bar{e}_L \mu_R + 
(2m_e - m_\mu - m_\tau) \bar{\mu}_L \tau_R + (2m_\mu - m_\tau - m_e) \bar{\tau}_L 
e_R] + \nonumber \\ 
&& {\phi_2^0 \over 3 v}[(2m_\mu - m_\tau - m_e) \bar{e}_L \tau_R + 
(2m_\tau - m_e - m_\mu) \bar{\mu}_L e_R + (2m_e - m_\mu - m_\tau) \bar{\tau}_L 
\mu_R] + H.c., \nonumber \\
\end{eqnarray}
and those to quarks are
\begin{eqnarray}
&& {\phi_1^0 \over 3 v}[(2m_b - m_d - m_s) \bar{d}_L s_R + 
(2m_d - m_s - m_b) \bar{s}_L b_R + (2m_s - m_b - m_d) \bar{b}_L 
d_R] + \nonumber \\ 
&& {\phi_2^0 \over 3 v}[(2m_s - m_b - m_d) \bar{d}_L b_R + 
(2m_b - m_d - m_s) \bar{s}_L d_R + (2m_d - m_s - m_b) \bar{b}_L 
s_R] + \nonumber \\ 
&& {\bar{\phi}_2^0 \over 3 v}[(2m_t - m_u - m_c) \bar{u}_L c_R + 
(2m_u - m_c - m_t) \bar{c}_L t_R + (2m_c - m_t - m_u) \bar{t}_L 
u_R] + \nonumber \\
&& {\bar{\phi}_1^0 \over 3 v}[(2m_c - m_t - m_u) \bar{u}_L t_R + 
(2m_t - m_u - m_c) \bar{c}_L u_R + (2m_u - m_c - m_t) \bar{t}_L 
c_R] + H.c. \nonumber \\ 
\end{eqnarray}
Contrary to Eqs.~(8) and (9), where many of the Yukawa couplings are 
proportional to the masses of light quarks and leptons, all the couplings 
in Eqs.~(13) and (14) are proportional to the largest mass in each sector, i.e. 
$m_\tau$, $m_b$, and $m_t$.  From the nonobservation of the rare decays 
listed in Eqs.~(3) to (6), this would require $m_{1,2}$ to be much greater 
than the constraints coming from Eqs.~(8) and (9).

\noindent \underline{\it Phenomenological Constraints}~:~
The best experimental limit on the rare decays listed in Eqs.~(3) to (6) 
comes from $K_L^0 \to \mu^\pm e^\mp$ with a branching fraction~\cite{pdg08} 
less than $ 4.7 \times 10^{-12}$.  Using Eqs.~(8) and (9), with $m_K = 495$ 
MeV, $m_s = 104$ MeV, $V_{us}=0.225$, the ratio
\begin{equation}
{\Gamma(K_L^0 \to \mu^\pm e^\mp) \over \Gamma(K^+ \to \mu^+ \nu)} 
= {9 m_K^2 m_s^2 \over 4 |V_{us}|^2} \left( {1 \over m_1^2} + 
{1 \over m_2^2} \right)^2,
\end{equation}
leads to the bound
\begin{equation}
{m_1 m_2 \over \sqrt{m_1^2 + m_2^2}} > 510~{\rm GeV}.
\end{equation}

Using the above bound, the branching fraction of the rare decay 
$B^0 \to \tau^+ e^-$ is then predicted by Eqs.~(8) and (9) to be less 
than $1.4 \times 10^{-7}$, well below the current experimental bound 
of $1.1 \times 10^{-4}$.  The other flavor-changing leptonic decays of 
$B^0$ (i.e. $\mu^+ \tau^-$ and $e^+ \mu^-$) are suppressed even further, 
i.e. by $m_\mu^2/m_\tau^2$ and $m_e^2/m_\tau^2$ respectively.  Similarly, 
the branching fraction of $B^0_s \to \tau^+ \mu^-$ is predicted to be 
roughly equal to that of $B^0 \to \tau^+ e^-$, and the other modes 
$\mu^+ e^-$ and $e^+ \tau^-$ are suppressed analogously.  The decay rates 
of $\tau^+ \to \mu^+ \mu^+ e^-$ and $\tau^+ \to e^+ e^+ \mu^-$ are 
proportional to $m_\mu^2 m_\tau^2 (m_1^2 + m_2^2)^2/m_1^4 m_2^4$ and 
$m_e^2 m_\tau^2 (m_1^2 + m_2^2)^2/m_1^4 m_2^4$ respectively and thus many 
orders of magnitude below current limits.  Similarly, the decay rates 
of $D^0 \to \tau^- e^+$ and $D^0 \to e^- \mu^+$ are proportional to 
$m_\tau^2 m_c^2 (m_1^2 + m_2^2)^2/m_1^4 m_2^4$ and $m_e^2 m_c^2 
(m_1^2 + m_2^2)^2/m_1^4 m_2^4$ respectively, and are also negligible. 

\noindent \underline{\it Higgs Decays}~:~
If the neutral Higgs bosons $\psi^0_{1,2}$ of Eq.~(10) are observed.  
Their decays are then completely determined according to Eq.~(8) and (9). 
They should decay dominantly into $t \bar{c}$ and $t \bar{u}$.

In the version with Eqs.~(11) to (14), $\phi_{1,2}$ are presumably too 
heavy to be observed.  However, $\eta$ and $\phi_0$ may be light, 
and their decays according to Eqs.~(11) and (12) would also be indicative 
of {\it flavor triality}.

\noindent \underline{\it Lepton Flavor Triality Alone}~:~
The notion of {\it flavor triality} is much more applicable to charged 
leptons than quarks, because neutrinos are nearly massless.  If quarks are 
not considered, then only the decays of Eq.~(3) are relevant and 
only Eqs.~(8) and (13) are to be studied.  Using Eq.~(8),
\begin{equation}
B(\tau^+ \to \mu^+ \mu^+ e^-) = {9 m_\tau^2 m_\mu^2 (m_1^2 + m_2^2)^2 \over 
m_1^4 m_2^4} B(\tau \to \mu \nu \nu) < 2.3 \times 10^{-8},
\end{equation}
the bound
\begin{equation}
{m_1 m_2 \over \sqrt{m_1^2 + m_2^2}} > 39~{\rm GeV}
\end{equation}
is obtained instead of Eq.~(16).  The branching fraction 
$B(\tau^+ \to e^+ e^+ \mu^-)$ is further 
suppressed by $m_e^2/m_\mu^2$ in this case.  On the other hand, the two 
branching fractions are about the same if Eq.~(13) is used, yielding 
a bound of 54 GeV instead.  This means that $\psi^0_{1,2}$ may be light 
enough to be accessible at the Large Hadron Collider (LHC). Their 
production~\cite{cmr07} is presumably via $Z \to \psi^0_{1,2} 
\bar{\psi}^0_{1,2}$.  Using Eq.~(8), $\psi^0_{1,2}$ are predicted to 
decay equally into $\tau^+ \mu^-$ and $\tau^- e^+$, whereas the $\mu^+ e^-$ 
mode is suppressed by $m_\mu^2/m_\tau^2$.  Using Eq.~(13), the decay rates 
of $\psi^0_{1,2}$ into $\tau^+ \mu^-$ and $\tau^- e^+$ are again roughly 
equal, but that to $\mu^+ e^-$ is now about four times larger.

\noindent \underline{\it Conclusion}~:~
The notion of {\it flavor triality} for quarks and leptons may a useful 
guide for checking if $Z_3$ is an underlying residual symmetry for 
understanding neutrino tribimaximal mixing from either $A_4$ 
or some other non-Abelian diecrete symmetry which also breaks to $Z_3$. 
In two specific scenarios with scalar mediators, flavor-changing quark 
and lepton interactions are discussed.  If only {\it lepton flavor triality} 
is valid, these scalars may be light enough to be observable at the LHC 
and their decays would reveal their underlying flavor structure.

\noindent \underline{\it Acknowledgement}~:~ 
This work was supported in part by the U.~S.~Department of Energy under
Grant No. DE-FG03-94ER40837.

\baselineskip 16pt
\bibliographystyle{unsrt}

\end{document}